# Associating eHealth Policies and National Data Privacy Regulations

Saurav K. Aryal
Computer Science
Howard University
Washington, DC
saurav.aryal@howard.edu

Peter A. Keiller
Computer Science
Howard University
Washington, DC
pk@scs.howard.edu

*Abstract*—As electronic data becomes the lifeline of modern society, privacy concerns increase. These concerns are reflected by the European Union's enactment of the General Data Protection Regulation (GDPR), one of the most comprehensive and robust privacy regulations globally. This project aims to evaluate and highlight associations between eHealth systems' policies and personal data privacy regulations. Using bias-corrected Cramer's V and Thiel's U tests, we found weak and zero associations between e-health systems' rules and protections for data privacy. A simple decision tree model is trained, which validates the association scores obtained. *(Abstract)*

*Keywords—Data Privacy, e-Health systems, e-Health policies, WHO dataset, nominal association (keywords)*

## I. Introduction

With the ubiquity of electronic data in modern society, privacy concerns increase. Nations and privacy regulations highlight this concern. The United States started the trend back in 1996 with its enactment of the Health Insurance Portability and Accountability Act (HIPAA). The European Union enacted legislation that required even more robust data privacy controls in its General Data Protection Regulation (GDPR). Furthermore, even China's Cryptography Law provides some protection for data privacy. Concurrently, as privacy regulations adapt to this hyperconnected modern world, there is widespread digitization of health records, even in developing countries. The World Health Organization's data [1] shows that 46% of nations have national e-health record (EHR) system regulation.

Consequently, important questions arise: is there an association between a countries' EHR system regulations and national data privacy laws? Do countries with EHR regulations already have personal privacy laws in place? Do specific EHR regulations better associate with national privacy laws than others?

### A. Contribution

This research aimed to determine if a statistical association exists between eHealth regulations and privacy laws globally. To the best of our knowledge, we are not aware of any research seeking to quantify the association between these variables statistically. We also create and share a new dataset containing these variables.

### B. Organization

The following section describes the dataset and methodology used for our analysis. We then follow with the results from our research. Finally, we end with a discussion, conclusion, and acknowledgments.

## II. Methodology

This section outlines the methodology, creating the dataset, and the statistical analysis techniques used for this paper.

### A. Dataset

No dataset in the existing research literature contains the variables of interest for this research. However, we found two independent datasets of interest to create a dataset for this project's purpose.

First, the World Health Organization's Global Health Observatory data repository [1] contains reported eHealth policy for 125 countries. The features included in this dataset are renamed and detailed in Table 1. In [1], the "Yes" responses to 1, "No" responses were converted to 0, and "Don't know" or "No response" responses were converted to -1. However, the data does not contain variables that reflect each of the nations' privacy laws.

Second, The United Nations Conference on Trade and Development (UNCTAD) has a dataset [2] of the name of data or privacy legislations for each country. Two columns were created: national data privacy and national data privacy. If the legislation title contained the word "personal" or "data" or any known variants of the terms, the countries were labeled 1 in the respective columns. In countries like Barbados, the legislation name and title did not explicitly indicate whether the regulation involved policies related to either personal or data law. In those cases, the legislation in question was read to evaluate if the emphasis was placed on personal or data privacy.

TABLE I. Dataset Columns from WHO Dataset

| Original column label | Renamed Column |
|---|---|
| The country has policy or legislation to define medical jurisdiction, liability, or reimbursement of eHealth services such as telehealth | eHealth liability |
| The country has policy or legislation to address patient safety and quality of care based on data quality, data transmission standards, or clinical competency criteria | eHealth data quality |
| The country has policy or legislation to protect the privacy of individuals' personally identifiable data irrespective of whether it is paper or digital format. | PII protection any format |
| The country has policy or legislation to protect the privacy of individuals' health-related data held in electronic format as an EHR. | Individual Health Privacy |
| The country has a policy or legislation that governs digital data sharing between health professionals in other health services in the same country through an EHR. | Portability domestic |
| The country has a policy or legislation that governs digital data sharing between health professionals in health services in other countries through an EHR. | Portability International |
| The country has a policy or legislation that allows for sharing of personal and health data between research entities. | Portability Research |
| The country has a policy or legislation that allows individuals to access their own health-related data when held in an EHR. | Individual access |
| The country has a policy or legislation which allows individuals to demand that their own data be corrected when held in an EHR if it is known to be inaccurate. | Correction on demand |
| The country has policy or legislation which allows individuals to demand the deletion of health-related data from their EHR | deletion on demand |

Countries that did not meet these requirements were assigned 0 on the specific columns.

After cleaning and relabeling, both datasets were merged into a single dataset. We retrieved ISO 3166-1 (official) country names connected with regional codes from [3] and added a region column to the final dataset to foster granular research. The final dataset contains 123 countries with a total of 13 nominal columns. However, for this project, we only utilized data points obtained from [1] and the national data privacy laws column as obtained from [2].

*B. Statistical Association*

Correlation is a commonly used measure of the statistical association between variables. Pearson's Correlation Coefficient (PCC) is the standard metric used. However, PCC is used for linear relationships and is not defined for nominal data. The typical approach to using PCC with nominal data includes one-hot encoding of the features. However, this approach causes the feature space and complexity to increase.

There are statistical methods that aim to solve this problem. These methods can be broadly classified into distance metrics and contingency table analysis. Out of these two, distance metrics, while intuitive, are scale-dependent. This property is undesirable when evaluating the association between different features. Moreover, standard conversion of a distance metric into a coefficient of association may not exist. Although contingency table analysis suffers from fewer drawbacks than distance metrics, the correct analytical method's choice is critical. There are numerous coefficients, which use the chi-square statistic: Phi coefficient [4], Contingency coefficient C [5], Tschuprow's T [6], Cramer's V [7], and Theil's U [8]. Of these, the Phi coefficient is only defined for 2x2 tables. Both Contingency coefficient C and Tschuprow's T are better suited for associations involving a greater number of classes and do not mathematically reach the maximum value of 1.0. Thus, we chose Cramer's V and Theil's U test. Both tests score the association in the range [0, 1], where 0 means no association and 1 is complete association.

The original formulation of Cramer's V, as derived from PCC, is known to suffer from bias and tends to overestimate the strength of association between variables [9]. We opt to use the bias-corrected version of Cramer's V provided by [9]. Despite bias correction, Cramer's V assumes a symmetrical correlation between the variables. For any two nominal variables, $a$ and $b$, utilizing Cramer's V for quantifying the association between them. The association score $V$ has the property outlined in (1).

$$V(a) = V(b) \quad (1)$$

While (1) can be a simplifying assumption, the symmetrical assumption may cause information loss. As such, an asymmetric measure is required. To preserve and analyze any asymmetric measure of association between our variables, we choose to utilize Theil's U. While [8] returns scores between [0, 1] similar to [9], the association score $U$ between two nominal variables, $a$ and $b$, is not symmetrical as defined in (2).

$$U(a) \neq U(b) \quad (2)$$

Both [8] and [9] will be utilized on the data points gathered from [1] and the national data privacy law column from [2] to evaluate the statistical strength of association between these variables. As of now, the remaining data points are not evaluated. A matrix of association scores will be generated to examine the associations between all of the variables associated using both measures. The difference in results obtained between [8] and [9] will be evaluated.

*C. Decision Tree Model*

A Decision Trees is a non-parametric supervised learning method used for classification and regression tasks. Decision trees learn from data to approximate a sine curve with a set of if-then-else decision rules. A simple decision tree classifier based on the Gini criterion is used to predict the binary labels for national data privacy using the features obtained from [1]. We expect the performance metrics of prediction to be in line with the statistical association measures utilized. To get a better idea of the performance, we use k-fold cross-validation accuracy across these folds is reported. We purposefully chose a higher train size to capture as much of the variation as possible in the dataset. To reduce the bias in the accuracy scores produced, we decide to use k-fold cross-validation with k=10. As this is a binary classification task, we created a confusion matrix and defined accuracy in (3), where A is Accuracy, TP is True Positives, TN is True Negatives out of N test samples.

$$A = (TF + TN) / N \quad (3)$$

*D. Programmatic Analysis*

All of the data preprocessing, gathering, and analysis is performed programmatically. We chose to use the Python [10] programming language as it was well-supported, and we were most familiar with it. Well-known open-source libraries are used. Seaborn [11] and matplotlib [12] are utilized to create figures and plots for this research. Scientific computing and data science libraries: NumPy [13], SciPy [14], Dython [15], Pandas [16], and Sci-kit Learn [17] are used for data-preprocessing, file I/O, statistical analysis, and decision tree modeling as outlined in the preceding sections.

III. RESULTS

The association scores were calculated for each pair of variables from our dataset using both [8] and [9]. The resulting matrices are shown in Fig. 1. The rows are labeled on the left, whereas each column is separately labeled for each matrix.

The results from [9] indicate weak association scores from 0.14 to 0.35 with the existence of national data privacy laws. Unsurprisingly, a higher association was noted between data points obtained from [1] with a maximum of 0.70. However, these scores are likely elevated with the assumption of symmetricity inherent in [7]. Interestingly, the highest association of national data privacy was observed with domestic data portability with a score of 0.35. The lowest scores were with liability in eHealth and correction of eHealth data on demands with scores of 0.19 and 0.14, respectively. The full score matrix can be seen on the left of Fig. 1.

Without the assumptions of [7], the association scores using [8] were remarkably lower for each possible pair. The association scores with national data privacy laws range between 0.02 to 0.08, i.e., no significant association. None of the columns from [1] had a noticeably higher or lower association with national data privacy. Nonetheless, similar to Cramer's V, higher association scores were observed between data obtained from [1] with a maximum score of 0.51 between individual access and correction on demand. These scores are on the right matrix in Fig. 1.

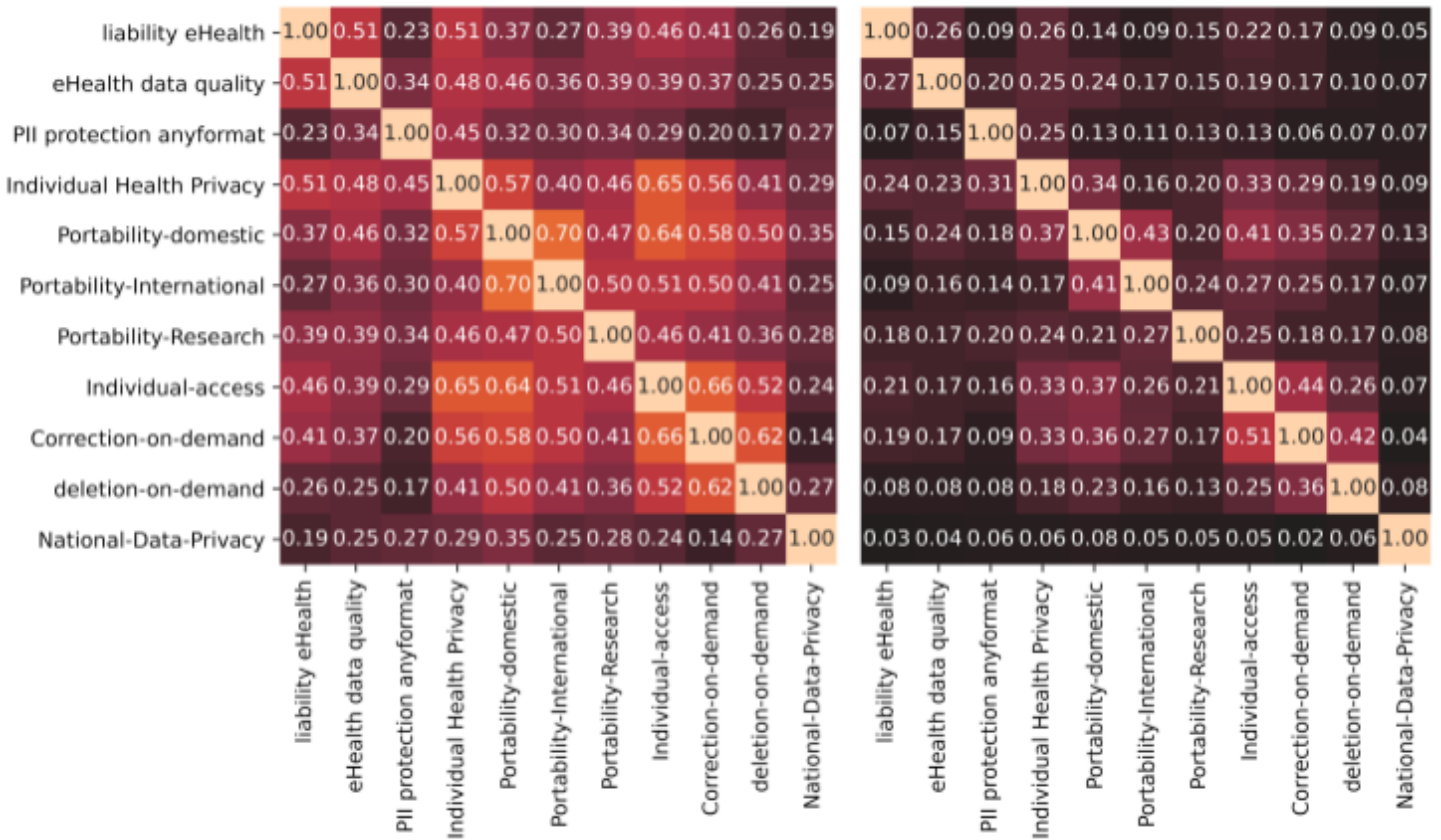

Fig. 1. Association scores with Cramer's V (left) and Theil's U (right)

The decision tree classifier reported an average accuracy of 0.69 across the ten folds. This accuracy is slightly higher than that by random chance for binary classification. However, given the marginal probability of national data privacy, as illustrated in Fig. 2, the classifier could not find any associations as expected from our statistical test results above.

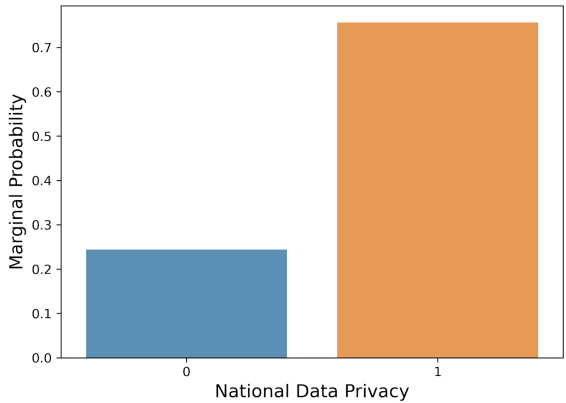

Fig. 2. Marginal Probability Distribution of National Data Privacy.

IV. DISCUSSION

As expected, the decision tree model underperformed compared to the marginal probability distribution of the national data privacy data column. This reaffirms our findings of a weak to an insignificant association between the features and the target variable.

Without the context of the individual countries' regions, the weak association is troubling, specifically in cases where policies and practices govern electronic health data without the requisite data privacy laws in place. Similarly, while there might be strong data privacy laws in place, countries can yet utilize them and expand the access and policies related to electronic health data. Specifically, the lack of association between individual health privacy, PII protection, liability in eHealth with national data privacy laws is most troublesome as these are usually most significant to personal health data.

While regional analysis would have provided a more granular perspective from the dataset, this analysis urges countries and governments to reconsider their regulations governing personal data privacy in light of the recent surge of eHealth technologies and policies. However, geopolitical, socioeconomic, and cultural differences between countries might also explain the lack of statistical associations between these variables. Regardless, an overhaul at the global level overtime is needed to strengthen data privacy protections associated with the increasing demand for eHealth technologies and policies that govern such systems.

V. FUTURE WORK

A region-specific analysis is the next step of this research where regional associations can be drawn concerning eHealth policies and data privacy laws. Personal privacy laws may also be added as an additional feature for further analysis.

As of now, the dataset is limited to eHealth policies and privacy legislation. No human factors such as cultural, economic, or financial data are considered. These factors can be added to expand the dataset to enable the evaluation of associations between them and the data points that are currently provided.

The dataset can be used in health systems or to expand existing research in the area. To aid in this, we will be making the dataset publicly available along with the source code for this project on a public GitHub repository https://github.com/github-user/repo-for-project.

VI. CONCLUSION

Using Cramer's V and Theil's U tests, we found a low to an insignificant association between eHealth policies and existing data privacy laws across 123 countries. These findings were further corroborated by training a decision tree classifier, resulting in a low average accuracy of 0.69, which is worse than achievable through marginal probability analysis across 10-fold cross-validation. This lack of association is troubling as it implies that while countries may have policies in place for eHealth capabilities, these capabilities are not necessarily associated with existing national data privacy laws to protect citizens' data.

The dataset and the source code for the project will be made openly available for further research in this area.

ACKNOWLEDGMENT

While he was unable to be included in this paper's authorship due to his current research obligations and guidelines, we, the authors, would like to acknowledge the help, guidance, and support received from Mr. FirstName LastName. Mr. LastName's help was critical in the ideation of this project and the creation of this project. With the authors, the dataset was created as a class project for graduate coursework at Name College. We would also like to acknowledge the help received from Dr. Firstname Lastname to guide us through the statistical analysis required for this project. This project would not have been possible without both of your assistance.